# Towards Secure IoT: Securing Messages Dissemination in Intelligent Traffic Systems

**Jawdat Alshaer[1]**

[1] *Department of Computer Information Systems, AL-Balqa Applied University, AL-Salt, Jordan*

**Abstract:** A few years ago, Automotive area in the IoT was seen as theoretical concept and today we are already seeing the possibilities of not only driverless cars, but applications of IoT in the intelligent vehicles including parking, maintaining environment, protecting lives and smoothing the flow vehicle movements. We have realized the urgent need of using simple and efficient secure protocol in Vehicular Ad Hoc Network (VANET) to be practical in the fast mobility of the network nodes, and taking advantage of the existence of base stations gateways along the road to inherit the protocol to different VANETs, this will reduce the initialization of communication overhead time and the security keys initialization each time a node passes to new base station zone. In this research, we applied security protocol used in sensor networks to achieve security in VANET, the simulation analysis shows that secure practical communication is achieved which can be inherited to other sub VANETs. The contribution of this article is enhancing proposed protocols with as less cryptography computation overhead as possible to make it applicable in the high mobility nature of VANET using security primitives; which guarantees security while allowing fast authenticating during vehicle passing one VANET to the next one depending on its direction in the transportation networks.

## 1. INTRODUCTION

Internet of Things (IoT) is a new internet technology wave which is transforming human lives. The traditional internet, which used to connect people, now has been extended to connect things. This will provide communication to intelligent service pooling, using low cost internet connected devices and sensors while applying intelligent algorithms. In these days, people lives and jobs require them continuously to be both mobile and online, even when they are onboard driving a vehicle. The diagnostics OBD/OBD-II port, (which is like a computer) monitors emissions, mileage, speed, and engine functionalities, besides road traffic and conditions. This information can be displayed on special screens or sent to service specialists for analysis. Alerts related to the vehicle like open doors, lights on and hand brake, or related to the trip conditions require driver or automated vehicle control to perform actions on certain vehicle parts such as Lock/Unlock vehicle doors, start the ABS system and in cases to safely stop the vehicle. Vehicles also equipped with GPS, Gyroscopes Orientation and Accelerometer Sensors, which can be used to model the driving behavior.

Safety Sensors are the basis of safety systems and focus on recognizing accident hazards and events almost in real-time. Micro-mechanical oscillators, speed sensors, cameras, radars and laser beams, inertial sensors, ultrasonic sensors, proximity sensors, night vision sensors [1].

By mounting the specific equipment in the vehicle, sensors readings can be processed to match driving patterns with road conditions, such as icy road, turns, obstacles and traffic jams. Monitoring different parameters actively while processing and intelligently analyzing sensors readings, drivers can get real time alerts and warnings on their screens and smart devices and technical faults can be identified for early handling. By allowing Vehicular Ad hoc Networks (VANETs) between vehicles, Roadside Units (RSU), and Base Station (BS) as in figure 1, some useful information regarding road status and emergency situations can be readily made available to other vehicles, government authorities, and service stations; this will enable a better traffic performance and leads to lives saving.

*E-mail :jawdat_alshaer@bau.edu.jo*

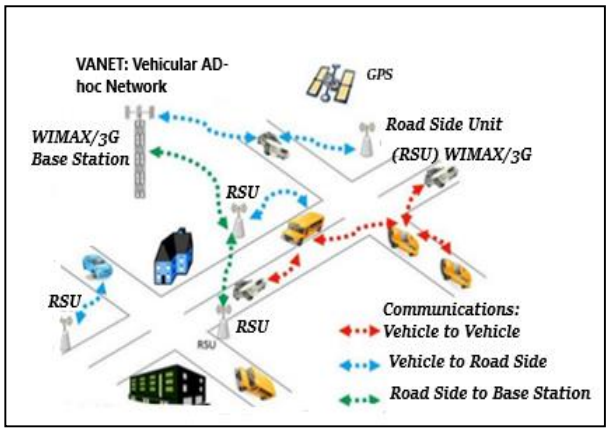

Figure. 1. Vehicular Ad Hoc Networks in transportation system.

Vehicle data communication is new added network technology as a result of the revolution in vehicle technical and communication specifications. Each moving vehicle equipped with sensors, processing units and transmission channels, with location and context aware properties [2]. Base Stations (BS) on roads form the gateways for these networks. Vehicles enter signal range of one BS, joining one VANET then exit to another, allowing communication between same zone vehicles and the BS. VANET deploys and integrates different technologies for simple, effective and secure communication: Wi-Fi IEEE 802.11, WAVE IEEE 1609, WiMAX, IEEE 802.16, Bluetooth, IRA, and ZigBee. The communication can be extended to rich media streaming communication between vehicles, infotainment and telemetric. In the high mobility of VANET, mobile IPv6 is an enhanced version of Mobile IP; this enhancement allows messages exchange in VANET to be performed even during fast continuous location changes. VANET can take advantage of all these communicating capabilities intelligently applying the right connection in the right spatial and temporal properties. However, assuring security; as in all other new technologies; still a challenge and VANET is not an exception.

Vehicular Ad hoc Network (VANET) was derived from Mobile Ad hoc Network (MANET) which has a self-organizing architecture, it used to connect mobile devices via wireless link, in this infrastructure there is no centralize authority, which means both of the networks are self-organized and decentralized systems. Security approaches must not rely on central services provided by fixed infrastructure as nodes in these networks are independently mobile. Due to the absence of the centralize authority, the attacker easily can join the network and perform any malicious activities. Various trust based security architectures were proposed to overcome this fault. This research proposes applying security primitives of sensor networks on VANET protocols in order to maintain data communication in secure manner in the high mobility nature of VANET.

## 2. Proposed Intelligent Secure Transportation System Network Architecture

The purpose of this research is to achieve Secure Intelligent Transportation System (SITS). Layered structure of VANET; as in figure 2; gives the ability to isolate the main functionality from the basic infrastructure, and allows intelligent algorithms to process and classify different sensors data readings, these algorithms to be embedded in the application layer.

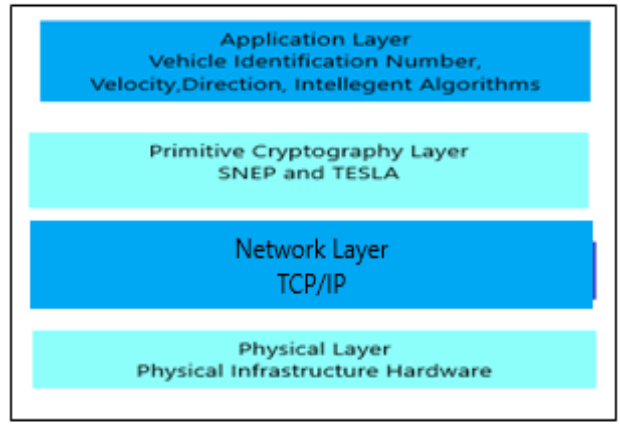

Figure 2. Layered Vanet structure to apply security primitives

As part of IoT; achieving SITS requires intelligent technical inspection and classification of vehicles sensors data [3] and [4], allowing messages exchange, securing the sources, and applying intelligent choice of cryptography. Forcing semantic security and achieving the principle of categorizing who needs to know and which information in real critical time and in a securely and urgently fashion. However other non-critical messages still can be transmitted using any of wide complicated proposed protocols, the intelligent classifying of messages is essential in this process.

Intelligence gives VANET the best quality of service as it was introduced in [5], to smoothly manage traffic flows, improving the traffic experience, and appropriately reduce the life loss and risks in the transportation roads. The deployment of intelligent road side gateways which usually have high processing power, big memory and bandwidth allows running specific classifying algorithms and visualization of information with video and audio support. This will reduce the VANET processing and loads handling on the mobile bodies.

Applying the proposed layered architecture, figure 2, will require constructing VANET structure with intelligent layers for implementing pool of classified services and functionality and flow of messages. This research proposes security primitives to be part of VANET layer structure and to be applied in the transmission of sensitive classified sensors data in fast secure manner.

## 3. Literature Review

Applying cryptography on VANET communications using symmetric keys is more efficient in computation and transmission terms, however, it doesn't protect from the repudiation of messages. Using digital signatures were proposed to solve this problem, such as using Group Signatures (GS) in [6]; where table of signatures of all nodes are being saved and maintained in each node to prevent repudiation of messages. In [7], the traditional Public Key Cryptography (PKC) was proposed to achieve the semantic security. In this work, the reliability of messages totally dependent on the number of nodes delivering the same information. Vehicles allowed frequently to change their identities. This approach can consume large storage and cause delay of messages transmission. Using identity-based signature & batch verification was proposed in [8]. This security paradigm approach requires updating the anonymous identity in time synchronized manner. However, almost all proposed works require buffering of each node identical signature and continuously changing identity in the network and the overhead of maintaining and transmission of keys between nodes, In this current work, symmetric mechanisms are proposed while assuring integrity through the use of message counters, and time synchronized key disclosure which was originally proposed for sensor networks in [9]. Raw et al. in [10] highlighted different security mechanisms for security routing protocol and integrity, which can guide the researchers for more efficient integrity in VANET, still, the computation and transmission cost of applying these mechanisms is high. Multi Channels Model proposed by Shukla et al. in [11]. For securing VANET network functionality while assuring confidentiality. However, it cannot be forced on industries infrastructure to be used in all vehicles types. Islam et al. in [12] used Public Key mechanism to assure authentication, integrity, and non-repudiation security requirements in VANET, the methods used includes updating security certificates when vehicle enters new VANET zone decreasing authentication time and by result, decreasing the message delay time.

Ming-Chin et al in [13], proposed privacy preservation authentication scheme (PPAS) for VANETs. PPAS intended to be light weight and to solve many security issues, but can be applied to vehicle to base stations messages exchange only, without considering the vehicle to vehicle communication security issues. In earlier research, Ming-Chin et al. [14] realized the need for security protocol for VANET to be practical in the high mobility environment nature of VANET, they propose trust-extended authentication mechanism (TEAM) for VANET with lightweight cryptography, they performed a simulation proving the efficiency of the proposed security methods. However, their work partly relies on technical hardware modifications which can be considered a restriction of their work.

There are many security methods proposed for VANET, a complete list of proposed VANET security and privacy methods were well studied by Marvy B. et al. in [15] and summarized in table 1.

TABLE 1. Summary of VANET Security and Privacy Approaches

| Security and Privacy Requirements | Security and Privacy Methods |
|---|---|
| Authentication, Privacy | Credential Usage, Digital Signature, Encryption, Anonymizer Proxy |
| Authentication, Data Integrity | Credential Usage, Digital Signature, Encryption, Message Authentication Code (MAC) |
| Authentication, Data Integrity | Credential Usage, Digital Signature, Encryption, Message Authentication Code (MAC) |
| Anonymity, Unsinkability | Pseudonym Usage, Silent Period, Mix-zone |
| Traceability, Accountability, | Credential Usage, Digital Signature |

Most proposed security protocols and methods assure security but impractical in VANET; as each node should buffer all other nodes private keys and certificates, which is difficult and can result in storage overloading, besides transmission delay. To achieve the semantic security A. Perrig et al. in [9] proposed two secure protocols (SNEP and TESLA) for securing messages in sensor networks, where SNEP guarantees confidently, authentication, integrity and freshness. TESLA is used for authenticated broadcasting and strong freshness assurance. They recommended their security protocols for higher level of communications, the current research is adopting their methods for sensors data transmission in VANET. We propose that sensor protocols can be applied on emergency messages when classified urgent intelligently, by an algorithm resides in base stations, choosing the right security protocol for the right packet to assure secure fast delivery for urgent sensors readings messages. However other entertainment messages and streams can use different protocols to allow rich contents communication. In this work we focus on transmitting short messages using well studied sensor security primitives proposed and proved efficient in [9].

## 4. Technical network Specifications used in VANET

VANET applies modified version of IEEE 802.11p, which was developed by IEEE for vehicular networks. The modifications were mostly done in the network layer, figure 2. There are many wireless technologies like IEEE 802.11p that is a standard for Dedicated Short Range Communication, DSRC [16], including:

- Wifi type called Wireless Access in Vehicular Environment, WAVE,
- General Packet Radio Service (GPRS),
- IEEE 802.16 that is a standard for WiMAX,
- 4G-Long Term Evolution (LTE) which have been proposed for reliable vehicular communications,
- Modified version named 802.11p was developed by IEEE for vehicular networks, for dynamic multimedia support Wifi IEEE 802.11 and WiMAX IEEE 802.16 are used.

VANET allows communication between vehicles and roadside stations, according to its spatial temporal parameters. This provides the direct flow of messages between vehicles with or without the side stations gateways, Mobile IPv6 proxy based architecture supports these two scenarios. VANET is intended to connect vehicles with each other and with fixed stations (mobile and stationary nodes) which provides gate way to WAN. Vehicles equipped with GPS, communication and processors are location aware as stated in [1] and get connected with each other and to fixed road gateway locally, by periodically exchanging messages contain their identity, time, velocity, direction and empty content. List of active nodes is maintained in the ad hoc network, where inactive ones are deleted.

### A. Layered communication Architecture

The previously explained layered architecture of vehicle nodes provides different capabilities and constraints as the followings:

- It can deploy a different set of applications including intelligent ones,
- The nodes usually moving at high speeds (0-40 m/s),
- There are no constraints on battery and storage,
- VANET consists of vehicle nodes and Base station which is the gateway to wide area network,
- Nodes are moving routing forest while the BS is the root of the tree, means each VANET can be represented with a tree data structure. VANETs are a forest (multiple trees),
- The periodic transmission of messages (beacons) constructs the routing topology where the addressing allows the exchange of messages.

### B. VANET Security Requirements.

- Trust Requirements: Usually, VANETs are distributed in different zones, that is vehicles can move continuously passing untrusted zones. This can securely threaten the network communications. Butting in mind the fact that wireless communication is untrusted broadcast, and different eavesdrop, replay and injection of messages by adversary are expected, hence, the proposed security protocol must force security on each node including the BS, adopting this strategy from [9], all vehicle nodes must trust the BS initially through a key shared between both then sequence of keys are derived from the original one. Each node has local clock which is used for the broadcast protocol.

- Data Confidentiality: Vehicle nodes must exchange only encrypted messages using secured channels only.

- Data Authentication: Vehicle nodes must be authenticated before the information is adopted in the receiver side. Data authentication is achieved through symmetric key for Message Authenticating Code (MAC) calculation. Despite using MAC, any receiver knows the key can send malicious messages with faking identity, solving this problem, requires asymmetric mechanism generated from the symmetric one using the contribution of [9], by constructing asymmetric broadcast using delayed key disclosure and one-way function

- Data Integrity ensures that messages were not altered, in this work integrity is already deployed by data authentication policies

### 5. PROPOSED SECURITY PROTOCOL

To achieve the required security, two building blocks were adopted from [9], namely, the design and the implementation of two secure protocols SNEP and TESLA for securing messages, where SNEP guarantees confidently, authentication, integrity and freshness. Secret key is shared between each node and the BS then between the sender and receiver nodes. The protocol extended to establish trust between all network new nodes. As we propose dividing the VANET in an area into group of sub-VANETs where each is connected with individual base station gateway. TESLA is used for authenticated broadcasting and strong freshness assurance. Table 2 lists the notations and descriptions which will be used in constructing the proposed security protocol.

TABLE 2. PROPOSED PROTOCOL NOTATION AND DESCRIPTION

| Notation | Description |
|---|---|
| A, B | Communication nodes |
| NA | The nonce, can be bit string, generated by A |
| M1\|M2 | Concatenation of message M1 and message M2 |
| KAB | Symmetric shared key between A and B |
| Km | Secret master key between base station and node |
| CAB | The cipher text of message M generated using KAB |
| C(KAB, x) | Cipher text with block chaining |
| {D} | Encrypted Data |
| CTR | Implicit counter of messages in a node |

### A. Security Enforcement using SNEP

It was proved in [9], that applying SNEP in sensor networks, will assure Confidentiality, Authentication, Integrity, and Freshness, the cryptography primitives were efficiently used in sensors network with restricted memory, processing and energy.

We claim that these advantages by applying SNEP in VANET are possible and effective as the similar main restriction in VANET is lowering the communication bandwidth and delay time between vehicles while moving fast passing from one VANET to another, as SNEP works simply by adding bytes to messages, providing semantic security while assuring efficient communication protocol. The magic in that, is the appended bit string to the encrypted chain of messages with block chain DES algorithm for example. For achieving two nodes authentication, MAC code is used.

The SNEP provides semantic security without transmission overhead by sharing counter between nodes. Node implicit counter is incremented while messages are being exchanged, and doesn't need to be transmitted. The sequence of messages will be differently encrypted according to each node implicit counter, to assure freshness a nonce can be sent from requested node and the responder one will send back the nonce as part of the response message. The cipher text is generated by applying the encryption key and the counter on the original message [9], as in (1):

$$Cipher\ Text = \{D\}<Kenc, CTR> \quad (1)$$

Where *D* is the data encrypted using the encryption key Kenc, concatenated with the implicit counter (CTR) in the sender. Message Authenticated Code (MAC) to identify the sender to the receiver node is generated, the CBC-MAC from [16] is used and concatenated with the cipher text as in (2):

$$M = MAC\ (Kmac, CTR\ |\ C) \quad (2)$$

All keys are derived from master key using one-way function as in equation (5).

The complete message sent from A to B is constructed from equations (1) and (2) as in [9], including the two parts: encrypted message along with MAC, as in (3):

$$A \rightarrow B: \{D\}(Kenc, CTR), MAC(Kmac, CTR\ |\ C) \quad (3)$$

The previous encryption protocol, assures the following advantages [9]:

- Low Communication requirements as the counter is incremented implicitly and no need to be transmitted while the vehicle is moving in the range of current VANET where the node belongs; this will increase security while lowering the transmission time which is critical in VANET topology and it is the core goal for this research.

- Data authentication: whenever a message received, the MAC must be verified true.

- Replay protection: The synchronized implicit counter in both sides will prevent replaying old messages.

- Right sequence of messages between nodes is assured by each node implicit counter.

- Data freshness is achieved by nodes using nonce which is generated and sent with requests and the nonce is be sent back within the response message that is being embedded in the MAC computations in the response message, response message may contain shaking hands information, general or emergency ones, as in (4)

$$A \rightarrow B : Na, Ra$$

$$B \rightarrow A :$$

$$\{R_B\}(Kenc, CTR),\ MAC(KMAC, NA\ |\ CTR\ |\{R_B\}(Kenc, CTR\ )) \quad (4)$$

Node *A* sends a nonce to node *B* using random number generation, Node *B* responds with secure encrypted nonce message, then *B* receives and verifies the MAC, recognizes the responder node and the contents.

### B. Authenticated Broadcast using TESLA

Most literature proposals for authentication in VANET are using asymmetric digital signatures which requires long time of communication (delay) for creating the signature, sending, receiving and verification overhead before starting information messages exchange. However, applying asymmetric mechanisms in VANET is impractical as the communication time is critical where vehicles passing fast between different base stations. To overcome this problem, TESLA protocol was proposed for providing broadcast while assuring security in [17] and [9]. TESLA can be applied in VANET as processing and energy is not limited, however, delay is critical and transmitting of messages must be in real time. Modified version to reflect faster authentication and broadcasting in VANET can be even more applicable. The TESLA protocol was redesigned as follows:

Modified TESLA (µTESLA), Requires that BS and vehicles are loosely time synchronized, clock time for each must be part of the key generation. Communication initialized by sending(broadcasting) packet from the base station to all nodes periodically, calculating the MAC, the nodes start communication with the BS receiving MAC on the message combined with the secret key, the received node verifies that MAC key is known only by the BS according to the synchronized time stamp, and so verifying that the original message was not altered and so stored in the receiver buffer ,soon, the BS will broadcast the key to all nodes which can have the key verified and used for further authenticating the stored messages. The BS and all nodes maintain chain of MAC keys generated by one-way function according to time schedule depends on the passing time of vehicle in the VANET, for any node to communicate at any particular time, the sender node picks the last key Ki and applies the one-way function in [9], to generate the other keys (key chain), as in (5)

$$K_i = F(K_{i+1}) \qquad (5)$$

The advantage of the one-way key chain is that once the receiver has an authenticated key of the chain, subsequent keys of the chain can be used and authenticated.

For each node, to communicate, it needs to perform time synchronization to obtain the key from the BS, while leaving one Sub VANET and connecting to the next one, as all VANETs apply the same protocol with different parameters and key chains.

Figure 3 shows an example of μTESLA. Each key belongs to sequence of keys updated periodically with static real time stamp authenticating messages. The same key chain is used in different sub VANETs, where each base station represents sub VANET on specific cluster(zone), allowing the stream of messages to continue in that zone, as messages usually related to specific zone conditions. Later in time, new key is used in the zone. At any particular time, the previous key for previous messages can be calculated according to the BS broadcasted current one, as the key is timely synchronized between nodes. This is applicable as most messages are real time ones, however, Previous keys can be derived too for old messages.

## 6. SECURITY PROTOCOL IMPLEMENTATION

In VANET, constraints on vehicle movements and the fast speed results in fast joining and disjoining different zones of networks, where each zone is VANET with BS gateway, this leads to challenges in feasibility, security and efficiency. Successful network communication requires strong cryptography, despite these challenges in VANET topology. VANET hardware includes many sensors, strong processing units and active transmission of data. All nodes in VANET equipped with sensors, controllers, transmissions and processors, information is generated and exchanged. Random number generation is easily produced using specific MAC function with private key.

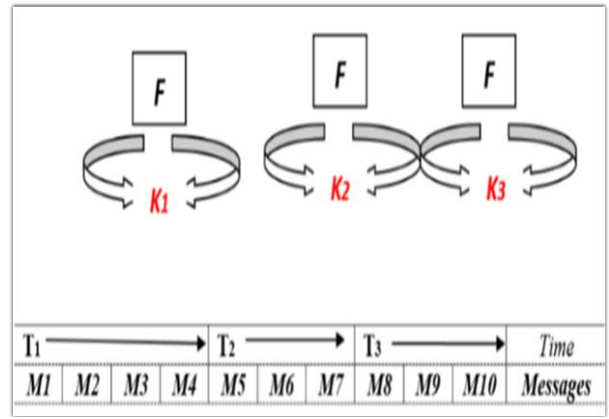

Figure 3. M1 and M2 are using k1 while later in time M9 and M10 will use the key K2

Choosing the right block cipher algorithm is critical, for example using the AES algorithm as in reference [16] with many patterns of bytes for lookup tables is too large to be processed in short time of node connection to VANET joining in fast speed. Good choice can be RC5 from [18] regarding its small code and high efficiency, it uses 32-bit data rotation and can be optimized for VANET with some reductions of code without affecting efficiency.

The counter may not be included in all messages; to keep the cipher text light in size in the VANET environment, as processing and transmission should be fast and in real time. Counter value is variable with time, which leads to different cipher for the same message in different time stamps this assures semantic security means repeated messages will generate different ciphers. To minimize messages weight and delay; the counter is not included in the messages transmission as it does exist implicitly in both sides the sender and the receiver nodes. New joining nodes can synchronize for using the same counter with BS using SNEP.

- Key setup is initialized in between a first node in VANET and the BS with secret master key, all followed chain of keys are derived from the original one, as shown in figure 3, M1 encrypted with both (key and counter) then xored with the current time function. This will give the advantage of continuous joining of the fast moving vehicle with all base stations while it is moving, as they apply the same key setup strategy.

- Message authentication: the block cipher in the protocol will be reused, then CBC-MAC can be used, block construction is shown in figure 4.

- For assuring message integrity, message M is encrypted with key $K_{enc}$, MAC key $K_{mac}$ as shown in (6):

$$\{M\}K_{enc}, MAC_{(K_{mac},\{M\}K_{enc})} \quad (6)$$

A MAC for each packet must be produced to handle the loosy nature of VANET communications, this MAC will guarantee integrity and authentication at the same time.

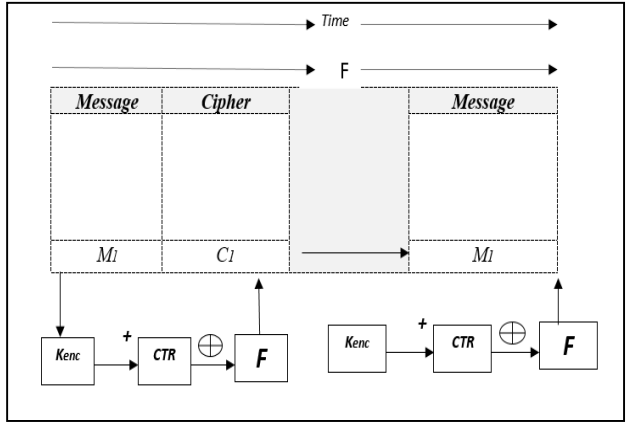

Figure 4. Cryptography primitives applied with the counter and time mode

## 7. PRIMITIVE PROTOCOL ANALYSIS

SNEP and µTESLA practically have been evaluated efficiently and used practically as stated in [9] for sensor networks, the transmission of messages requires only the size of 8 bytes for encryption and authentication, in VANET, where processing and buffering is not restricted as in sensor networks, different key generation algorithms can be applied. In particular, the semantic security in VANET can be efficiently achieved using simple techniques proposed for practical sensor networks.

For deeper study reflecting the efficiency and the simplicity of the proposed security primitives, simulations were conducted for location aware routing protocols namely DSDV, GPSR, and BMFR.

The Destination-Sequenced Distance-Vector (DSDV) Routing Algorithm, the nodes periodically transmit their routing tables to their neighbor nodes and any updates according to time synchronization or event-driven [19].

Greedy Perimeter Stateless Routing, GPSR, which utilizes the relation between geographic position of the node and message targeted nodes [20].

Neighbor node for forwarding message to destination node BMFR [21].

The parameters used in the NC2 Simulation were taken from [21]. Implementing our proposed protocol, the security size requirements are 40 Bytes for the RC5 module, 60 Bytes for µTESLA and 20 bytes for the encryption Module. The encrypted, small, message size plays basic rule in the transmission time, the storage usage in high traffic situations and the delivery ratio, as shown in the simulation results.

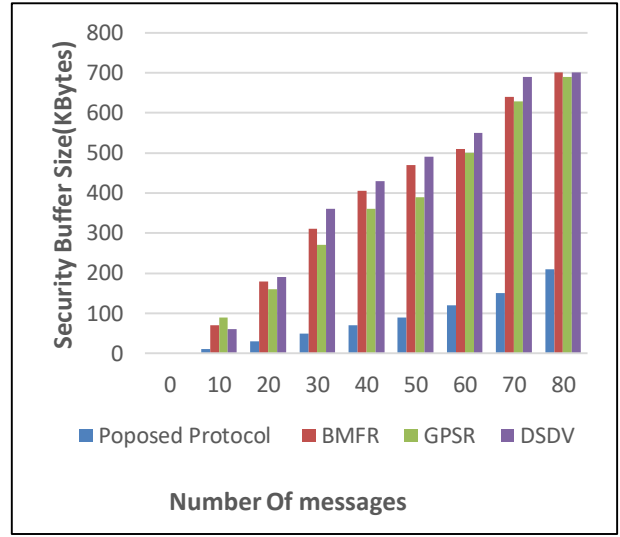

Figure 5. Storage usage vs. traffic load

Figure 5 shows that among all protocols, the proposed security buffer size is relatively very small, compared to the buffers required for applying the three routing protocols and that's logically true regarding the storage requirements for simple primitive cryptography applied on the proposed one which is initially has been used for sensor networks.

Delay of messages due to the security mechanisms where tested in the simulation, comparing to other routing protocols, as in figure 6.

The delay in the simulation, means the time taken for sent packet from node to reach the receiver node, including buffering, processing and transmitting. The delay of encrypted packets after applying the cryptography primitives, were very small and ignorable comparing to the other protocols, as the buffering, processing, and transmitting of the encrypted messages require low processing cycles and transmission time.

The proposed primitives were deployed for achieving this goal. Low delay is achieved by using stream cipher for encryption, where the size of encrypted message is almost the size of the plaintext. The MAC uses 8 bytes of every 30-byte message; however, the MAC also achieves integrity, that is, no need to use other message integrity mechanisms (e.g. a 16-bit CRC). Thus, encrypting and signing messages imposes an overhead of 6 bytes per message over an unencrypted message with integrity checking, or about 20 % of the overall size of packet [9].

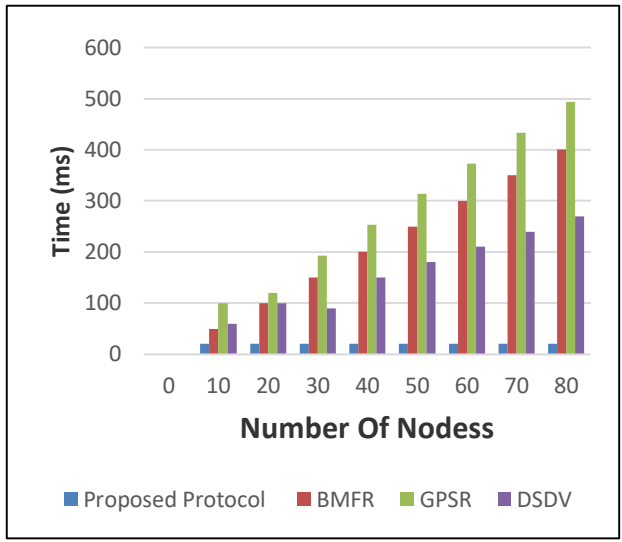

Figure 6. Average End to End delay of packets

The used algorithm RC5 in the proposed security protocol requires only 8 instructions per cycle which leads to very low processing and transmitting; encrypting 30-byte packet requires less than 4 cycles. Applying the security primitives proposed for sensor networks is adequate for the high mobility nature of VANET were the proposed methods can sign and encrypt each message even with single security round, the disclosure of key can be dynamically set to time intervals related to the density and mobility of VANET nodes.

## CONCLUSION

Exchanging messages in VANET in secure manner is critical issue and should be researched deeply by intelligently classifying the different types of messages according to the safety or entertainment categories. Safety and emergency messages produced by specific safety sensors reflecting emergency information must be secured and transmitted or broadcasted in fast manner with no delay or loss, for these types of messages, sensor security cryptography primitives are proposed in this article to be used, namely SNEP for authentication, encryption while assuring integrity and confidentiality. µTESLA for secure authenticated broadcasting. Simulation and analysis for the adopted protocols proved the efficiency of these primitives in securing messages with ignorable delay and minimum requirements of buffering and processing. This study brings up the idea of constructing sensor network in the core infrastructure of VANET, putting in the research area of VANET the driverless vehicles automated communications through secure sensor networks with intelligent capabilities.

Int. J. Com. Dig. Sys. 8, No.4, ..-.. (July-2019)

[16] U. S. National Institute of Standards and Technology (NIST). DES model of operation. Federal Information Processing Standards Publication 81 (FIPS PUB 81).

[17] Adrian Perrig, Ran Canetti, J.D. Tygar, and Dawn Song. Efficient authentication and signing of multicast streams over lossy channels. In IEEE Symposium on Security and Privacy, May 2000.

[18] R. L. Rivest. The RC5 encryption algorithm. *Proc. 1st Workshop on Fast Software Encryption*, pages 86–96, 1995.

[19] C. E. Perkins and P. Bhagwat, "Highly Dynamic Destination-Sequenced Distance-Vector (DSDV) for Mobile Computers," "Proc. ACM Conf. Communications Architectures and Protocols", London, UK, August 1994, pp. 234-244.

[20] B. Karp and H. T. Kung, "GPSR: greedy perimeter stateless routing for wireless networks," in Proceedings of the 6th Annual International Conference on Mobile Computing and Networking (MOBICOM '00), pp. 243–254, August 2000.

[21] Priyanka Goyal, Anish Soni, Ashu Dalal & Arun Jain, " A Review on Routing Protocols used in VANET", IJARCS Volume 5, No. 6, July-August 2014.

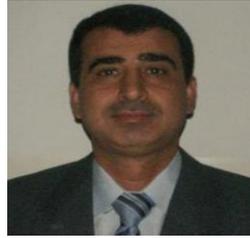

**Jawdat Jamil Alshaer:** Received the BSc degree from the Department of Computer Science, Mu'ta University, Jordan, in 1993, MSc degree from the Department of Computer Science, Wichita State University, USA, in 2003, and PhD. degree from the Department of Computer Science, Novosibirsk State Technical University, Russian Federation. He is currently Assistant Professor, working full time at Al-Balqa Applied University. Alshaer research interests cover topics from Spatial Temporal Databases, Networks, parallel Algorithms, Machine Learning, and Internet of Things.